\documentclass[fleqn,usenatbib]{mnras}

\usepackage{newtxtext,newtxmath}
\usepackage[T1]{fontenc}

\DeclareRobustCommand{\VAN}[3]{#2}
\let\VANthebibliography\thebibliography
\def\thebibliography{\DeclareRobustCommand{\VAN}[3]{##3}\VANthebibliography}

\usepackage{graphicx}	% Including figure files
\usepackage{amsmath}	% Advanced maths commands

\usepackage{comment}

\newcommand{\SI}[2]{#1\,#2}
\newcommand{\minute}{min}
\newcommand{\angstrom}{\AA}

\newcommand{\ML}[2]{%
  \if\relax\detokenize{#1}\relax
    \textcolor{olive}{\textbf{#2}}%
  \else
    \sout{#1}\,\textcolor{olive}{\textbf{#2}}%
  \fi
}

\title[Long-period LALOS in a solar filament]{Long-period longitudinal oscillations in a solar filament beyond the theoretical period limit}

\author[M. Rojas-Quesada et al.]{
Miguel Rojas-Quesada,$^{1,2}$
\thanks{E-mail: m.rojas-quesada.1@research.gla.ac.uk}
\href{https://orcid.org/0000-0002-3733-1714}{\includegraphics[scale=0.07]{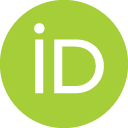}}
Manuel Luna$^{3,4},$
\href{https://orcid.org/0000-0002-3841-313X}{\includegraphics[scale=0.07]{img/ORCIDiD_icon.png}}
Lela Taliashvilli,$^{5}$
Heidy Gutierrez$^{5}$
\href{https://orcid.org/0009-0001-2490-2450}{\includegraphics[scale=0.07]{img/ORCIDiD_icon.png}}
\\
$^{1}$ School of Physics \& Astronomy, University of Glasgow, Glasgow G12 8QQ, UK \\
$^{2}$ School of Physics, Costa Rica Institute of Technology, Cartago, Costa Rica. \\
$^{3}$ Departament de Física, Universitat de les Illes Balears (UIB), E-07122 Palma de Mallorca, Spain\\
$^{4}$ Institute of Applied Computing \& Community Code (IAC3), UIB, E-07122 Palma de Mallorca, Spain\\
$^{5}$ Space Research Center (CINESPA), School of Physics, University of Costa Rica, San Pedro, Costa Rica
}

\date{Accepted XXX. Received YYY; in original form ZZZ}

\pubyear{\the\year{}}

\begin{document}
\label{firstpage}
\pagerange{\pageref{firstpage}--\pageref{lastpage}}
\maketitle

\begin{abstract}
Solar filaments exhibit oscillatory motions in different directions, including large-amplitude longitudinal oscillations (LALOs), which are commonly interpreted as magnetoacoustic oscillations with gravity providing the main restoring force In the standard pendulum model, the magnetic-field geometry imposes an upper limit on the oscillation period of approximately 167 min near the solar surface. Here, we report a longitudinal oscillation observed in a solar filament using SDO/AIA 193~\AA{} and 171~\AA{} observations, with periods ranging from 178--184 min. These values are longer than the near-surface cut-off period derived under the assumptions of the standard pendulum model. 
To investigate this discrepancy, we consider two possible effects: the prominence height above the solar surface and the three-dimensional geometry of the magnetic field. The former increases the theoretical cut-off period by reducing the local gravitational acceleration, while the latter can modify the longitudinal restoring force through the inclination of the plane containing the local field-line curvature. Specifically, a locally concave-down field-line geometry can reduce the restoring force and permit longer periods, provided that the equilibrium remains stable. However, the available observations do not allow us to independently constrain the prominence height, field-line curvature, and inclination. 
This result highlights the importance of considering the three-dimensional geometry and curvature of filaments when interpreting their longitudinal oscillations. Since LALOs provide a potential seismological diagnostic for estimating the magnetic-field strength supporting filaments, neglecting their curvature may introduce systematic errors in such estimates. Therefore, accounting for the geometry of the oscillating structure is essential for the accurate application of filament seismology.
\end{abstract}


%%%%%%%%%%%%%%%%%%%%%%%%%%%%%%%%%%%%%%%%%%%%%%%%%%

%%%%%%%%%%%%%%%%% BODY OF PAPER %%%%%%%%%%%%%%%%%%
\section{Introduction} \label{sec:intro}

Solar filaments, or prominences when observed at the limb, are relatively dense and cool plasma structures observed in the the corona.
At small spatial scales, they are composed of fine threads whose geometry, dynamics, and thermal properties are governed by the magnetic field. These structures are suspended against gravity by magnetic fields anchored in the photosphere \citep{Efremov2016}.

Filaments typically form above polarity inversion lines (PILs), where the radial component of the photospheric magnetic field changes sign, and the field becomes predominantly horizontal. Observationally, they appear as elongated dark ribbons located within, above, or far from active regions (ARs).

Large-amplitude longitudinal oscillations (LALOs) in solar filaments have been widely observed across multiple wavelengths since they were first reported by \cite{Jing2003PeriodicSubflare} (see also the review by \citealp{Arregui2018}).
Based on a theoretical model and hydrodynamic simulations, \cite{Luna2012} identified the projected gravity along the magnetic field as the main restoring force of LALOs, establishing the basis of the pendulum model adopted in this work. This result was subsequently
confirmed by \cite{Zhang2012} using one-dimensional radiative hydrodynamic simulations constrained by observations of a
longitudinally oscillating prominence. 

Although their simulated periods aligned well with observations, the damping times were overestimated by a factor of $\sim 1.5$--a discrepancy later attributed to mass drainage. 
Furthermore, \cite{Zhang2013} demonstrated that these oscillatory properties remain largely independent of the specific excitation mechanism (with variations below $1.4\%$). \cite{Luna2012b} showed that these oscillations correspond to the fundamental gravito-acoustic mode of the prominence, whose frequency is mainly determined by the geometry of the magnetic-field line through the projected gravity, with a smaller contribution from the slow-mode component associated with gas pressure. These results support the idea that the properties of LALOs are primarily determined by the filament magnetic geometry. 
LALOs provide a powerful diagnostic tool for probing prominence plasma and magnetic field properties through large-amplitude prominence seismology \citep{Luna2014,Luna2017,Kucera2022ComparisonChannel}. Understanding the factors that influence these oscillations is therefore essential.

The so-called pendulum model was extended by \cite{Luna2022ExtensionOscillations} to include the effects of non-uniform gravity. In this case, the oscillation period is given by
\begin{equation}
P = 2\pi \sqrt{\frac{1}{g_0 \left(\frac{1}{R} + \frac{1}{R_{\odot}}\right)}},
\label{eq:period_extended}
\end{equation}
where $R$ is the radius of curvature of the dipped magnetic field, $R_{\odot}$ is the solar radius, and $g_0$ is the gravitational acceleration at the solar surface. This expression was derived assuming that the magnetic-field dips
are located close to the solar surface and that the plane containing
the dipped magnetic-field line is perpendicular to the solar surface. These assumptions will be revisited in the present
work. 

In this model, the existence of magnetic dips is required for gravity to act as a restoring force. Under this assumption, \cite{Luna2022ExtensionOscillations} derived an upper limit for the oscillation period of approximately $P_{\odot} = \SI{167}{\minute}$, corresponding to the limiting case $R \rightarrow \infty$ in Equation ~\ref{eq:period_extended}. This value should be interpreted as the near-surface cut-off period within the assumptions of that model, rather than as an absolute upper limit for prominences at arbitrary heights or geometries.

In this paper we report the first observation of large-amplitude longitudinal oscillations with periods exceeding the theoretical upper limit proposed by \cite{Luna2022ExtensionOscillations}. We suggest that this discrepancy can be explained by the influence of inclination of the oscillation plane, considering that the previous models assume that the magnetic field lines lie in a plane perpendicular to the solar surface, ignoring the inclination of the oscillation plane.

This paper is organised as follows. Section~2 describes the methodology used to identify the LALO event in the filament under study and presents the observational results. Section~3 compares the observed oscillation with previous theoretical models, with particular emphasis on the role of filament morphology in explaining the measured oscillation period. Section~4 investigates the possible triggering mechanism of the event. Finally, Section~5 summarises the main conclusions of this study

\section{Methodology and Observational Results}

\subsection{Observational Data}
\leavevmode Filaments appear as absorption structures in EUV images. To track their motion and determine the oscillation period, we analysed images from the \textit{Atmospheric Imaging Assembly} (AIA) onboard the \textit{Solar Dynamics Observatory} (SDO) in the 193~\AA\ and 171~\AA\ channels \citep{Chamberlin2012}. AIA has a plate scale of 0.6 arcsec pixel$^{-1}$. We used a cadence of 3 min, while the oscillations considered here last from several minutes to a few hours. 

The SDO/AIA 193~\AA\ channel is primarily sensitive to coronal plasma, 
with peaks in its temperature response at $\log(T)=6.1$ and $7.3$, 
corresponding to $\mathrm{Fe\,XII}$ and $\mathrm{Fe\,XXIV}$ emission, 
respectively. The SDO/AIA 171~\AA\ channel has a peak response at a 
lower temperature, around $\log(T)=5.8$ \citep{Chamberlin2012}. In 
this passband, bright emission is often observed at the edges of the 
filament absorption region, which has been interpreted as emission 
from the prominence--corona transition region (PCTR) 
\citep{Kucera2022ComparisonChannel} and used to track filament material. 
The SDO/AIA 304~\AA\ channel, which is dominated by He\,II emission, 
is also used in this study to investigate the temporal evolution of the 
filament morphology on different dates, complementing the 
ground-based H$\alpha$ observations from GONG. This comparison is 
particularly useful when the filament becomes faint or partially 
disappears in H$\alpha$, as material may remain visible in the 
304~\AA\ images.

\begin{figure*}
    \centering
    \includegraphics[width=1\linewidth]{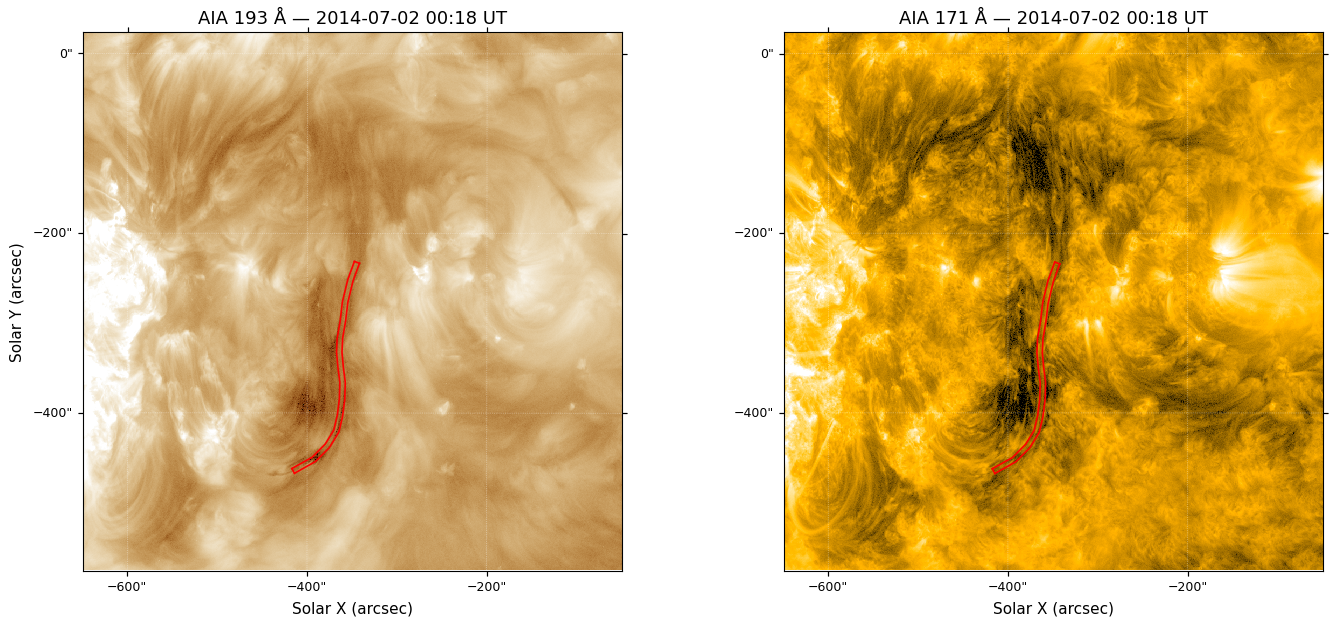}
    \caption{Curved slit used to analyse the filament oscillations. The filament, observed in the SDO/AIA 193~\AA\ and 171~\AA\ channel, is shown together with the finite-width curved slit of total width $W$ constructed along its axis. The slit follows the geometry of the filament and is used to sample and average the intensity across its width for the construction of the time–distance diagram.}
    \label{fig:img/F1}
\end{figure*}

The filament analysed in this study, with absolute coordinates S12°, 286° (CR2152), has an extent of $\approx$44° and is located 12.8° from active region AR 12104 (S14°, 269°). Additionally, another active region—AR 12107 (S19°, 256°)—is located to the south. The filament is oriented predominantly along the projected north–south direction in helioprojective coordinates extending across several longitudes (Figure \ref{fig:img/F1}). 

\begin{figure*}
    \centering
    \makebox[10pt]{
    \includegraphics[width=16cm]{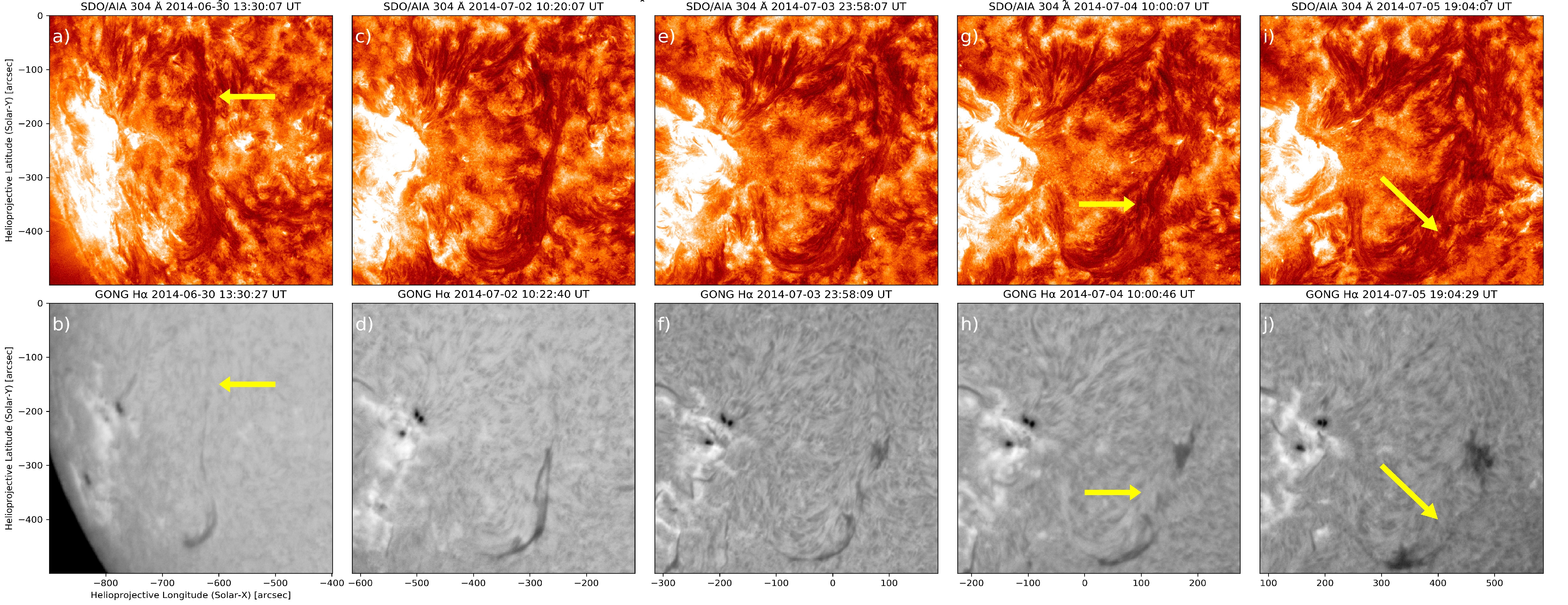}}
    \caption{Temporal evolution of the filament observed via GONG H$\alpha$ and SDO/AIA $\SI{304}{\angstrom}$ during the period from June 30, 2014, to July 5, 2014. The morphology is stable as of June 30, 2014, at 13:30 UT, although a weakening in the intensity of the northern section is observed in H$\alpha$ (a and b). On July 2, 2014, at 10:20 UT, the central and southern sections are fully visible in the H$\alpha$ images (c and d). The intensity of the condensed material decreases starting at 23:58 UT on July 3, 2014 (e and f). The central section disappears from the H$\alpha$ images (g and h) at 10:00 UT on July 4, 2014, while the presence of material is observed in SDO/AIA $\SI{304}{\angstrom}$. The material from the central section reappears in H$\alpha$ at 19:04 UT on July 5, 2014 (i and j).}
    \label{fig:TablaF2}
\end{figure*}

The temporal evolution of the filament from 30 June to 10 July 2014 shows a gradual change in morphology and intensity (Figure \ref{fig:TablaF2}). The filament remained relatively stable on 30 June, although the northern section weakened in H$\alpha$. On 2 July, the central and southern sections were clearly visible in H$\alpha$. From 3 July, the condensed material progressively decreased in intensity, and by 4 July the central section had disappeared in H$\alpha$, although material remained visible in SDO/AIA 304~\AA, indicating thermal instability in progress \citep{Taliashvili2009}. The central material reappeared in H$\alpha$ on 5 July. The H$\alpha$ observations were obtained from the Global Oscillation Network Group (GONG; \cite{Harvey1996}).

When the filament appeared above the limb as a prominence, a C5.0 flare associated with AR~12104 occurred on 9 July between 18:35 and 18:44~UT. On 10 July at 00:00~UT, the filament began a very slow total eruption (DBd), which ended at 06:40~UT. A spectacular CME became visible in LASCO~C2 from 07:12~UT, with a width of approximately $275^\circ$ and a position angle of $260^\circ$. According to the SOHO/LASCO CME catalogue, the CME had an average velocity of approximately 928~km~s$^{-1}$. SOHO/LASCO CME catalogue observations are described by \cite{Gopalswamy2009}.

The filament large amplitude longitudinal oscillations were examined using de-rotated data acquired on 2014 July 2, between 00:18 and 13:00 UT, when the filament morphology remains stable for the central and south sections, and no CMEs are detected.

\subsection{Construction of the Time--Distance Diagrams}

\leavevmode To construct the time--distance diagrams, we enhanced the contrast and sharpness of the SDO/AIA images using the wavelet-optimized whitening (WOW) process described by \cite{Auchre2023}. 
The movement of the plasma is tracked along a curved path defined by a slit which is specified by a set of manually selected control points, $(x_p, y_p)$, that are then linearly interpolated to generate a continuous sequence of coordinates, $(x_s, y_s)$, along the curved slit.

\leavevmode Following the procedure of \cite{Luna2018GONGMaximum}, we parameterized the central curve of the finite-width slit by its pixel coordinates $(x_s(s), y_s(s))$. The local tangent vector was estimated from finite differences along $s$.

From the tangent vector, we compute the corresponding unit normal vector $(n_x, n_y)$, defined as perpendicular to the local direction of the curve. This normal vector is used to sample points across the slit width.

A set of parallel offset curves is then generated by displacing the central curve along the normal direction by uniformly spaced distances $w \in [-W/2, W/2]$, where $W$ is the total slit width in pixels. The displaced coordinates are given by
\begin{equation}
(x_w, y_w) = (x_s + w n_x,\; y_s + w n_y).
\end{equation}

\begin{figure*}
    \centering
    \includegraphics[width=1\linewidth]{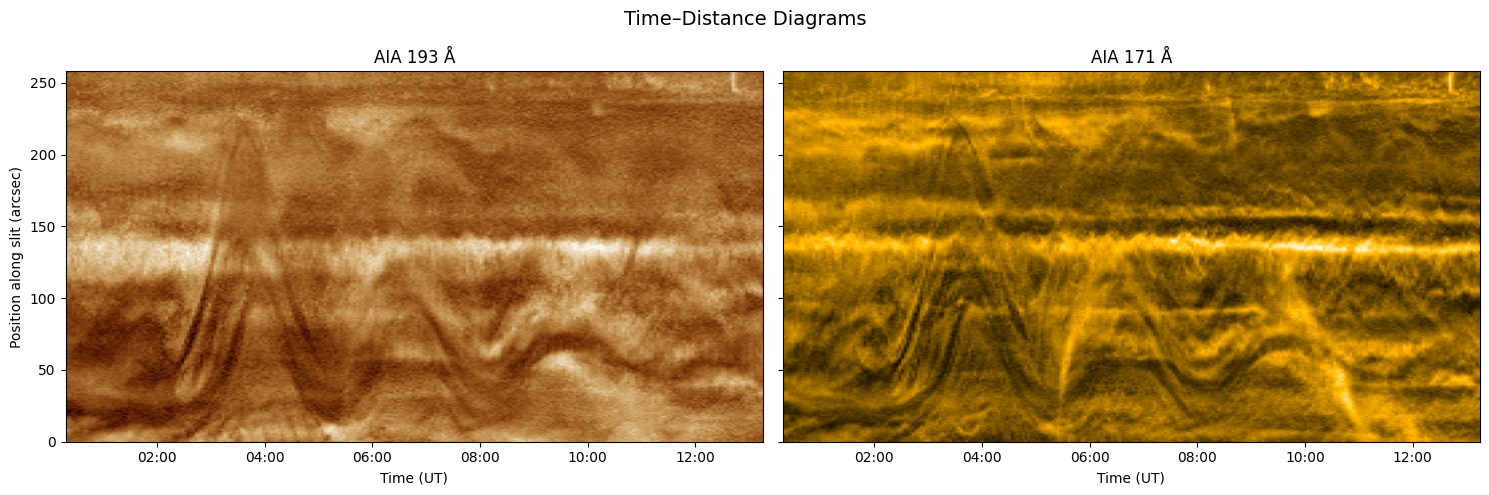}
    \caption{Time–distance diagrams of the filament oscillations. The diagrams show the temporal evolution of the intensity along the curved slit across the filament, as observed in the SDO/AIA 193~\AA\ and 171~\AA\ channels on 2 July 2014. The spatial coordinate represents the distance along the slit, while the horizontal axis corresponds to the observation time. The oscillatory motion of the filament material is manifested as alternating intensity structures propagating along the slit.}
    \label{fig:time_distance_diagrams}
\end{figure*}

The intensity is interpolated at each set of displaced coordinates using bilinear interpolation. The resulting intensity profiles are then averaged over all offsets to obtain a single representative slit profile. This same slit is applied to each frame of the derotated time series to construct the corresponding time-distance diagrams (see Figure \ref{fig:time_distance_diagrams}).

\subsection{Oscillation Characterisation}

From the time--distance diagrams, it becomes evident that tracking the filament oscillations is more robust when following the position along the slit where the filament plasma appears in absorption. Although emission, likely associated with the prominence--corona transition region, can also be identified at the edges of the absorption features and has been previously used to track filament oscillations \citep[see][]{Kucera2022ComparisonChannel}, this signal is difficult to track reliably because of the presence of background emission, which hinders the continuous identification of the oscillatory motion. 

\begin{figure*}
    \centering
    \includegraphics[width=1\linewidth]{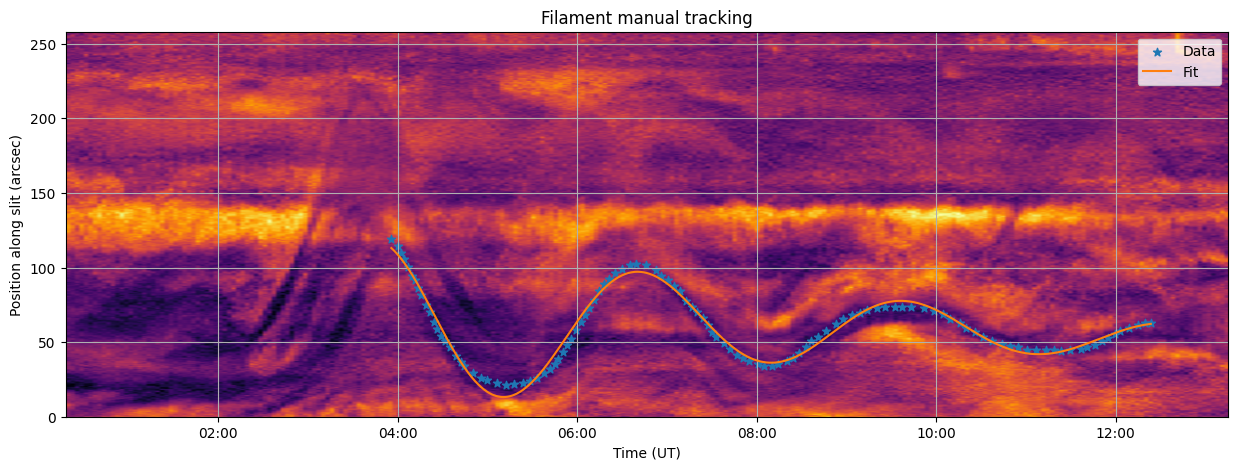}
    \caption{Manually selected filament positions ($s_c$) on the SDO/AIA 193~\AA\ time-distance diagram. The solid line represents the best fit to the tracked positions, using a damped sinusoidal function combined with a third-degree polynomial.}
    \label{fig:fit_sc_manual}
\end{figure*}

To track its position, two complementary methods are employed. First, a manual tracking is performed through visual inspection of the SDO/AIA 193~\AA\ time--distance diagram (Figure \ref{fig:fit_sc_manual}) between 03:55 UT and 12:23 UT on July 2. Second, an automated approach is applied by identifying local intensity minima along the slit at each time step and fitting them with a Gaussian function. This automated method is effective only when no additional absorption features are present within the same slit that could interfere with the tracking. Consequently, in our case, it is applied only to a segment of the time--distance diagram (between 6:30 UT and 12:45 UT on July 2) where the filament motion can be clearly distinguished from the background.

For the automated approach, we follow a process similar to that proposed by \cite{Luna2018GONGMaximum} to track oscillations in H-alpha. Under the assumption that  the filament absorption is determined by the column depth of cool plasma along the line of sight (LOS), the minimum intensity is taken to represent the central position of the structure along the slit. Accordingly, the filament is tracked over time by locating the intensity minimum at each time step.

\begin{figure}
    \centering
    \includegraphics[width=1\linewidth]{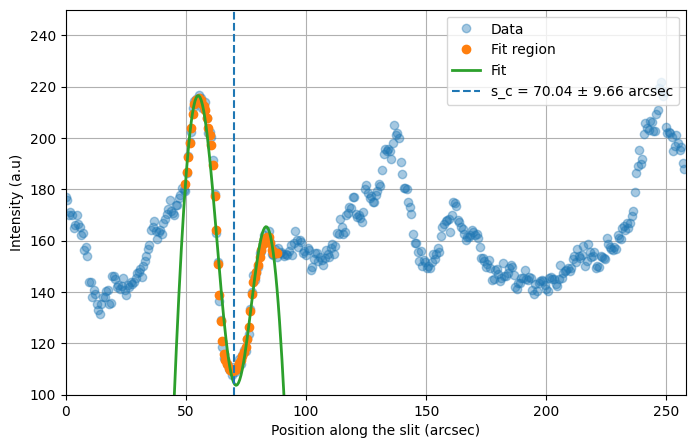}
    \caption{Intensity values along the curved slit at 2014-07-02 09:33 UT. The orange data shows the selected window for fitting, while the solid green line shows the fitted function. The filament centroid is estimated at position  $s_c = 70.04 \pm 9.66 $ arcsec.}
    \label{fig:fit_slit}
\end{figure}

\begin{figure*}
    \centering
    \includegraphics[width=1\linewidth]{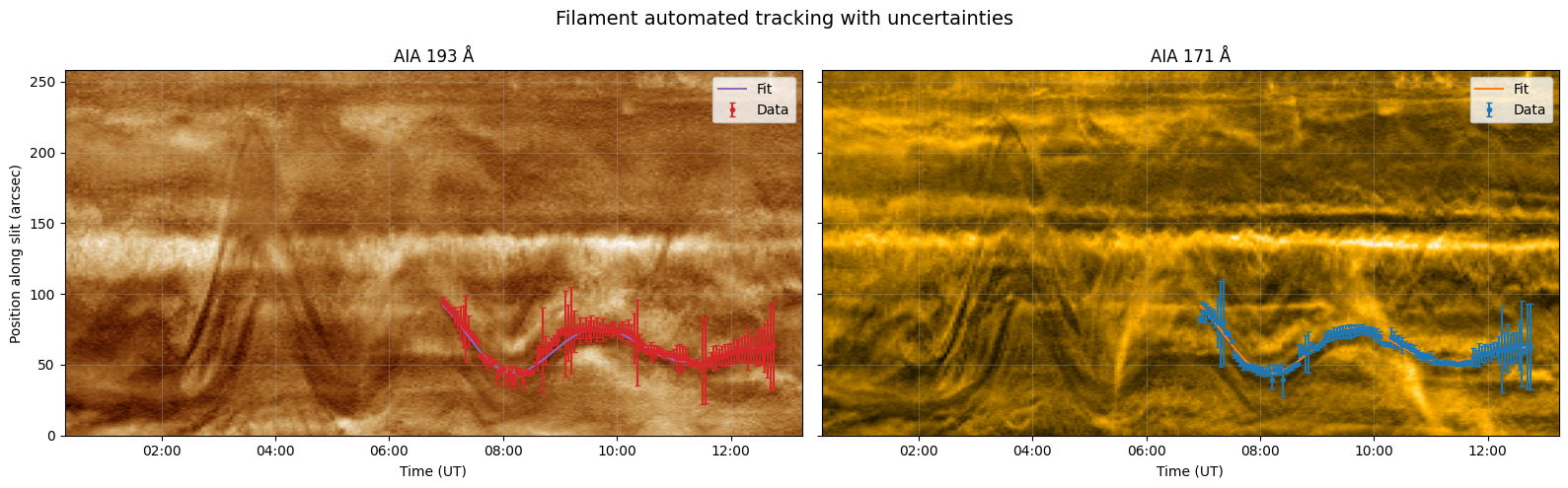}
    \caption{Values of the filament position ($s_c$) and their respective uncertainties (error bars) on the time-distance diagram, obtained using the automatic tracking method. The solid line represents the best fit to the tracked positions, using a damped sinusoidal function combined with a third-degree polynomial.}
    \label{fig:fit_sc}
\end{figure*}

Similarly, in our case the intensity profile for each slit is first smoothed using a Gaussian filter and a constrained search window centered on the previously detected position of the filament is used to locate the local minimum, ensuring temporal continuity in the tracking. Around this preliminary estimate, a narrower fitting window is defined, and the intensity profile is modeled using a Gaussian function with a background chromospheric emission that is fitted using a second-degree polynomial. The centroid of the fitted Gaussian provides an estimate of the filament position ($s_c$), while its width is adopted as a proxy for the positional uncertainty. An example of the described procedure for one slit is shown in Figure \ref{fig:fit_slit}. We repeated this procedure at each time step, updating the tracking position iteratively. Figure \ref{fig:fit_sc} shows the resulting estimates of $s_c$ and their uncertainties on the time--distance diagrams.

Following the approach of \cite{Luna2018GONGMaximum}, to determine the oscillation period and its associated uncertainty, the extracted signals were fitted with a damped sinusoidal function combined with a second-order polynomial term to account for long-term trends (see equation \ref{eq:fit_function}). The fitting and uncertainty estimation were performed using the Python routine \textit{Scifit}.
\begin{equation}
\begin{aligned}
s_c(t) ={}& A_0 e^{-A_1(t-t_0)}
\cos\left[A_2(t-t_0)+A_3\right] + A_4 \\
&+ A_5(t-t_0) + A_6(t-t_0)^2
\end{aligned}
\label{eq:fit_function}
\end{equation}
The fitting procedure was applied to both the automatically extracted and manually selected signals.For the automatically derived signal, the fitted Gaussian width was adopted as a proxy for positional uncertainty. For the manually selected signal, we assigned an uncertainty of 10 arcsec to each data point.

The fitting results yield oscillation periods of $178 \pm 2$ min for the manually selected signal, $184 \pm 3$ min for the automatically tracked SDO/AIA 193~\AA\ signal, and $183 \pm 3$ min for the SDO/AIA 171~\AA\ data. The close agreement between these measurements demonstrates the consistency of the two tracking approaches. These results show that, even when accounting for the measurement uncertainties, the observed period exceeds the near-surface cut-off period derived under the assumptions of \cite{Luna2022ExtensionOscillations}. The following section explores a possible explanation for this discrepancy.

\section{Implications for the Pendulum Model}
\label{sec:pendulum_model}

The observed oscillation
period, which lies between approximately 178 and 184 min for the
different tracking and fitting methods, exceeds the near-surface cut-off
period $P_{\odot}\simeq167$ min derived by
\cite{Luna2022ExtensionOscillations} under the assumptions of the
standard pendulum model. This discrepancy suggests that
some of these assumptions may need to be reconsidered for very
long-period LALOs. We first consider whether the gas-pressure gradient
can account for the observed period.

In general, the fundamental longitudinal mode is driven by the combined
action of projected gravity and gas-pressure gradients
\citep{Luna2012b}. Including the correction introduced by the curvature
of the solar surface, the period can be approximated as
\begin{equation}
\frac{1}{P^2}
=
\frac{g_0}{4\pi^2R}
+
\frac{1}{P_{\odot}^2}
+
\frac{1}{P_{\rm slow}^2},
\label{eq:period_pressure}
\end{equation}
where $P_{\rm slow}$ represents the contribution of the pressure-driven
slow mode. For the numerical model considered by
\cite{Luna2022ExtensionOscillations}, $P_{\rm slow}\simeq206$ min.

The gas-pressure gradient provides an additional restoring force and
therefore reduces the oscillation period. Its contribution becomes more
important for large radii of curvature, when the gravity-driven term is
small. Thus, although pressure gradients may be relevant for the event
analysed here, they cannot explain a period longer than the gravity-driven
cut-off.

The value of 167 min, however, was obtained under specific geometrical
assumptions. In particular, the prominence was assumed to be close to the
solar surface and the dipped magnetic-field line was considered in a
plane perpendicular to the solar surface. We consider below the effect
of relaxing these two assumptions. A more general theoretical study of
these configurations will be presented in a subsequent work.

\subsection{Effect of prominence height}

The cut-off period obtained by
\cite{Luna2022ExtensionOscillations} assumes that the prominence is
sufficiently close to the solar surface to take its heliocentric
distance as $R_{\odot}$. For a prominence at a height $h$, the
heliocentric distance is $r_0=R_{\odot}+h$ and the local gravitational
acceleration decreases as $g(r_0)=g_0(R_{\odot}/r_0)^2$. Following the
same argument as in \cite{Luna2022ExtensionOscillations}, the limiting
case $R\rightarrow\infty$ gives
\begin{equation}
P_{\max}
=
P_{\odot}
\left(\frac{r_0}{R_{\odot}}\right)^{3/2}.
\label{eq:pmax_height}
\end{equation}

Therefore, the 167 min value is not an absolute upper limit, but the cut-off corresponding to $r_0\simeq R_{\odot}$. For the observed range $P_{\max}=177$--$184$ min, Equation~(\ref{eq:pmax_height}) gives $r_0\simeq1.040\,R_{\odot}$--$1.067\,R_{\odot}$, corresponding to heights of approximately 28--46 Mm. A prominence located only a few tens of Mm above the solar surface could therefore have a gravity-driven cut-off period comparable to that measured in this event.

This value should be considered a lower limit. A finite radius of
curvature and the gas-pressure gradient both provide additional
restoring forces and reduce the oscillation period. Nevertheless,
prominence height alone can produce a significant increase of the
cut-off period and should be considered when studying very long-period
LALOs.

\subsection{Effect of the three-dimensional geometry}

\begin{figure*}
    \centering
    \label{fig:comparacion_planos}
    \begin{minipage}[b]{0.40\textwidth}
        \centering
        \includegraphics[width=\linewidth]
        {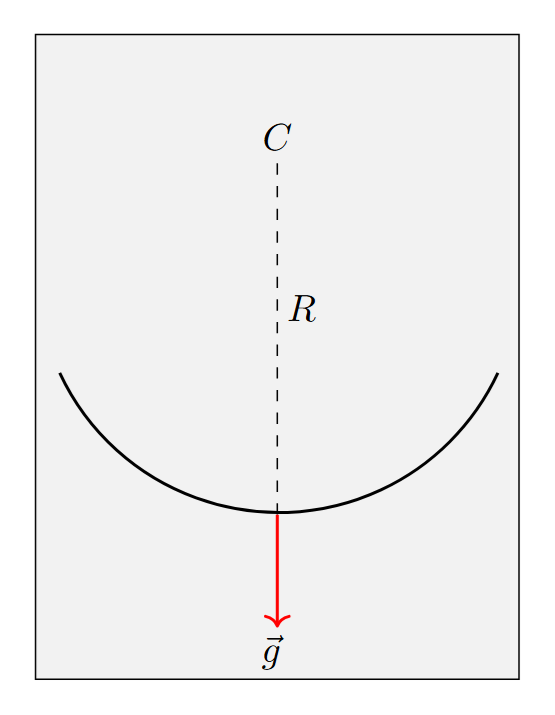}
        \par\smallskip\centering (a) Original 2D model.
        \label{fig:pla_osc_inic}
    \end{minipage}
    \hfill
    \begin{minipage}[b]{0.45\textwidth}
        \centering
        \includegraphics[width=\linewidth]
        {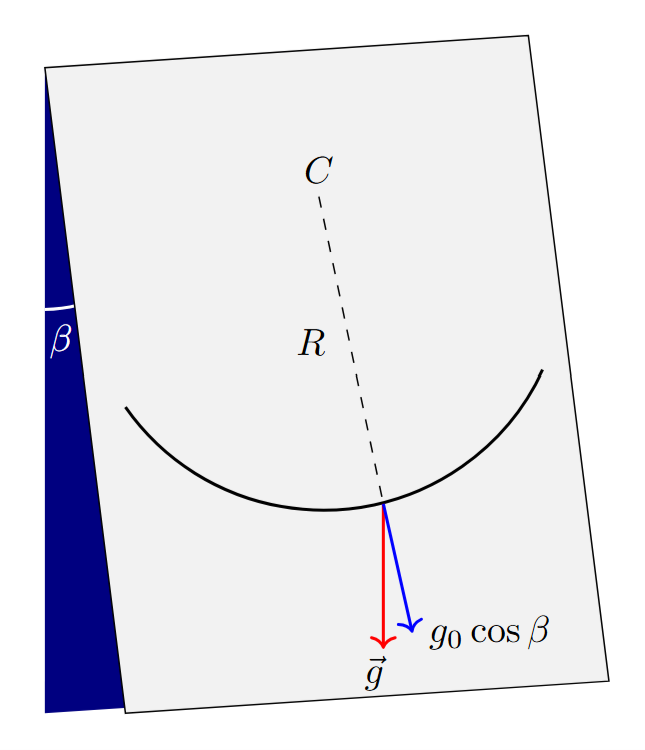}
        \par\smallskip\centering (b) Inclined plane of curvature.
        \label{fig:pla_osc_incl}
    \end{minipage}
    \caption{Schematic representation of the magnetic-field geometry. (a) Circular geometry considered by \protect\cite{Luna2022ExtensionOscillations}, where the plane containing the field line is perpendicular to the solar surface. (b) Geometry in which this plane is inclined by an angle $\beta$.  The inclination modifies the contribution associated with the local field-line curvature, but not the term associated with the curvature of the solar surface.}

\end{figure*}

A second possibility is related to the three-dimensional geometry of the
magnetic field. Figure~\ref{fig:comparacion_planos} shows the geometry
considered by \cite{Luna2022ExtensionOscillations} and a more general
configuration in which the plane containing the local curvature is
inclined by an angle $\beta$.

For this geometry, and allowing for a prominence at heliocentric
distance $r_0$, the gravity-driven frequency can be written as
\begin{equation}
\omega^2
=
g(r_0)
\left(
\frac{1}{r_0}
+
\frac{\cos\beta}{R}
\right),
\label{eq:omega_general}
\end{equation}
where $R$ is the signed radius of curvature of the magnetic-field line.
For $r_0\simeq R_{\odot}$ and $\beta=0$, this expression reduces to the
result of \cite{Luna2022ExtensionOscillations}.

An important consequence of Equation~(\ref{eq:omega_general}) is that
inclination alone cannot produce periods above the cut-off. For a
standard concave-up dip, $R>0$, the curvature term remains positive and
the period is always smaller than $P_{\max}$. However, the situation is
different when the signed curvature is negative. For $R=-|R|$, the
curvature term reduces the restoring force and periods larger than the
cut-off are possible while the equilibrium remains stable if
\begin{equation}
|R|>r_0\cos\beta.
\label{eq:stability}
\end{equation}

Thus, the combination of negative signed curvature and inclination
provides another possible explanation for periods longer than the
standard pendulum-model cut-off. A negative value of $R$ corresponds
to a locally concave-down field line, that is, to a loop-like geometry
rather than the concave-up dip usually considered in the pendulum model.
Interestingly, such a geometry can still be longitudinally stable when
the three-dimensional orientation of the field line is taken into
account.

For example, taking $r_0\simeq R_{\odot}$ and neglecting the
gas-pressure contribution, the observed period range of 178--184 min corresponds approximately to
\begin{equation}
|R| \simeq (5.7\! -\! 8.3) R_{\odot}\cos\beta .
\end{equation}
These values are only illustrative, since the present observations do
not allow us to determine $R$ and $\beta$ independently.

The two effects discussed above show that the observation of a period
longer than 167 min does not necessarily imply a failure of the pendulum
model. Instead, the result indicates that prominence height and the
three-dimensional geometry of the magnetic field become important for
very long-period oscillations. A positive-curvature dip could also explain the observed periods if located sufficiently high: height raises the cut-off, while inclination weakens the curvature contribution. For finite $R>0$ and $\beta<90^\circ$, the required height exceeds the $R\rightarrow\infty$ estimate and may be large. Gas pressure could increase it further. A detailed theoretical study of these effects will be presented in a subsequent work.

\section{Triggering mechanism of the longitudinal oscillations}

\begin{figure*}[H]
    \centering
    \includegraphics[width=1\linewidth]{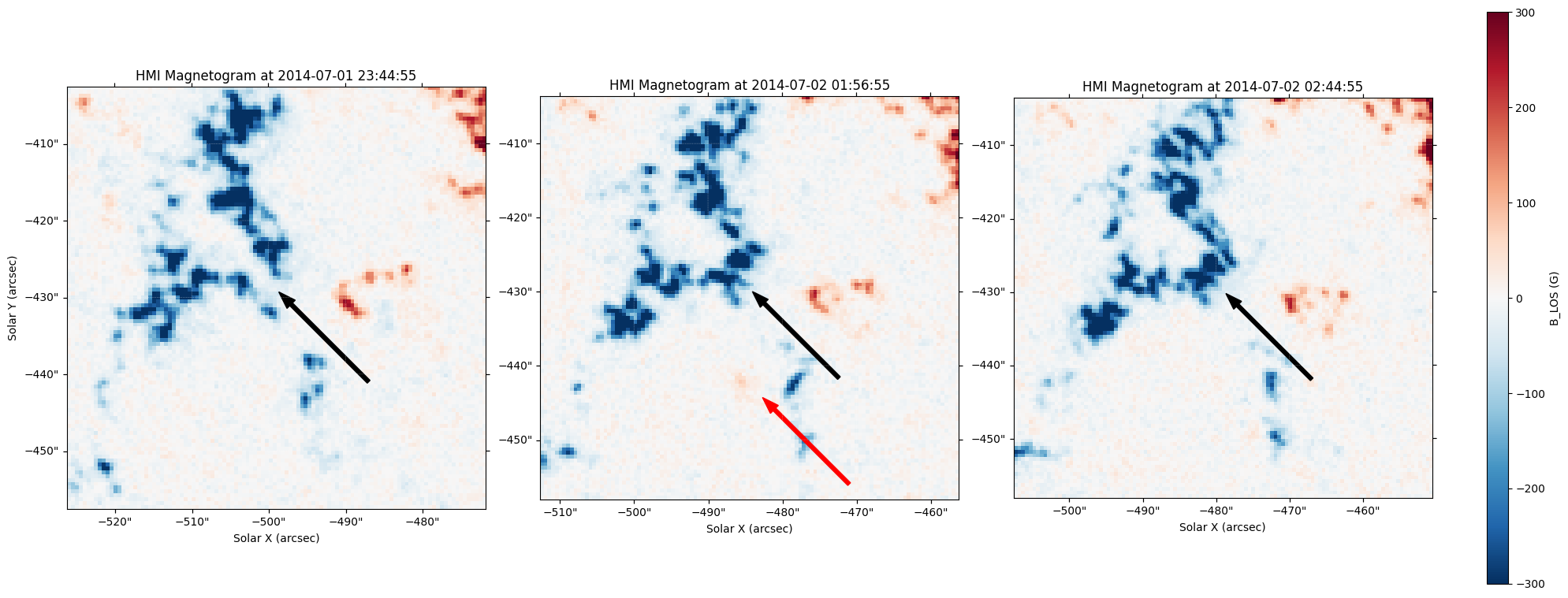}
    %\caption{Photospheric magnetic field evolution at the filament footpoint where the oscillations originate, as observed by HMI. The left panel shows the magnetic field before the oscillations begin, with two distinct negative-polarity monopolar regions separated from each other. The middle panel shows the corresponding photospheric magnetic field at the onset of the oscillations, when the two previously separated negative-polarity regions have merged. Coinciding with the merger, a small positive-polarity flux is observed to the south. The right panel shows the photospheric magnetic field one hour after the onset of the oscillations, when the positive-polarity flux has disappeared and the two negative-polarity regions have fully merged.}
    \caption{Photospheric magnetic field evolution at the filament footpoint, observed by HMI, before, during, and after the onset of the oscillations. The left panel shows the magnetic configuration two hours before the onset, characterized by two nearly separated negative-polarity regions connected by a weak magnetic field (black arrow). An extended negative-polarity feature approximately follows the projected trajectory of the oscillating filament material. This feature is absent from the daily Wilcox Solar Observatory magnetograms \citep{Scherrer1977}, which show a large-scale negative-polarity corridor extending towards both poles throughout the surrounding days, suggesting that the observed magnetic structure is a short-lived, localized process. The middle panel shows the onset of the oscillations, when the two negative-polarity regions merge and the magnetic field intensifies. This coincides with the emergence of a small positive-polarity flux to the south (red arrow). The right panel shows the magnetic configuration one hour after the onset.}
    \label{fig:hmi}
\end{figure*}

Reported triggers of prominence large-amplitude oscillations include Moreton/EIT/EUV waves and shock waves \citep{eto2002,shen2014a,shen2014b,gilbert2008}, nearby jets \citep{Luna2014,Zhang2017,joshi_interaction_2023}, subflares and flares \citep{Jing2003PeriodicSubflare,jing2006,Vrnak2007}, and filament eruptions \citep{isobe2006,Luna2017,mazumder_simultaneous_2020,Luna2024Excitation}. \citet{Luna2018GONGMaximum} identified a trigger in 85 of 196 catalogued oscillations: 72 were associated with flares, 11 with prominence eruptions, and one each with a jet and a Moreton wave. In one event, \citet{joshi_interaction_2023} observed magnetic-flux emergence near the jet base and a jet propagating along the filament channel, consistent with the model proposed by \citet{LunaMorenoInsertis2021Jet}.

Although no significant flare or CME was reported before the oscillations began at approximately 01:50~UT, the observations identify the filament footpoint from which they originate. We therefore examined the photospheric magnetic field near this footpoint during the preceding minutes. 
The oscillation onset coincides with the merger of two converging negative-polarity concentrations near the footpoint (Figure~\ref{fig:hmi}). The merger is accompanied by a plasma ejection toward the filament. This spatial and temporal association suggests that the interaction may have contributed to triggering the oscillations.

Interactions between small-scale photospheric flux concentrations, including same-polarity mergers, can restructure the magnetic field \citep{Ryutova2003InteractionElements}. \citet{Ryutova2003InteractionElements} discussed reconnection between non-collinear flux tubes, although reconnection between the two concentrations observed here cannot be confirmed. The merger coincided with a plasma ejection toward the filament and the appearance of a small positive-polarity concentration, which disappeared within an hour. This feature may indicate local magnetic reconfiguration, but its origin is uncertain. The observed sequence is consistent with a localized disturbance and plasma injection contributing to the onset of the oscillation.

\section{Conclusions}
In this work, we have analysed a LALO in a solar filament observed with the SDO/AIA 193~\AA\ channel on 2014 July 2. The main results of our analysis can be summarised as follows.

\begin{enumerate}

\item We identified a coherent longitudinal oscillation of the filament and measured its period using manual tracking and an automated method based on the centroid of the absorption feature in SDO/AIA 193~\AA\ and 171~\AA\ data. The manually selected signal yields a period of $178 \pm 2$~min, while the automatically tracked signals yield periods of $184 \pm 3$~min and $183 \pm 3$~min in the 193~\AA\ and 171~\AA\ channels, respectively.

\item The measured period is significantly longer than the maximum period of approximately 167~min predicted by the model of \cite{Luna2022ExtensionOscillations}, which considers gravity as the main restoring force and includes the effect of the curvature of the solar surface. The discrepancy cannot be resolved by including gas-pressure gradients, since pressure provides an additional restoring force and therefore acts to decrease, rather than increase, the oscillation period.

\item We discuss two possible explanations for the observed period range of 178--184~min. First, prominence height raises the cut-off. In the gravity-only limit, heights of approximately 30--46~Mm correspond to this period range. Finite positive curvature and gas pressure shorten the period, so greater heights may be required. Second, three-dimensional field geometry can modify the restoring force. Locally concave-down curvature may permit longer periods while preserving longitudinal stability. The observations do not constrain height, curvature, and inclination independently.

\item The geometrical correction has important implications for prominence seismology. Since the period of a LALO is commonly used to infer the radius of curvature of the supporting magnetic field and, consequently, to estimate the magnetic-field strength, neglecting the inclination of the oscillation plane may introduce systematic errors in these estimates. The observed period therefore suggests that the three-dimensional morphology of the filament should be considered when applying LALO-based seismological diagnostics.

\item We also investigated the possible origin of the oscillation. No significant flare or CME was observed immediately before the onset of the motion, while a clear temporal and spatial correspondence was found between the beginning of the oscillation and the convergence and subsequent merger of two same-polarity magnetic concentrations near one of the filament footpoints. The merger was accompanied by a localized plasma ejection toward the filament. These observations suggest that the magnetic-field interaction may have produced a localized disturbance or plasma injection that initiated the oscillation. However, the available observations do not allow us to establish magnetic reconnection as the specific mechanism responsible for the trigger.

The measured periods exceed the near-surface cut-off, and both prominence height and three-dimensional field geometry can modify the predicted period. The observations do not determine their contributions separately. Stereoscopic observations and improved magnetic-field measurements and extrapolations will be needed to test these interpretations and refine LALO-based seismology.

\end{enumerate}
\section*{Acknowledgements}

We acknowledge the Global Oscillation Network Group (GONG) for providing the H$\alpha$ observations used in this study. The data used in this work are courtesy of NASA/SDO and the AIA and HMI science teams. SDO data are managed and processed by the Joint Science Operations Center (JSOC) at Stanford University. The SOHO-LASCO CME catalog is generated and maintained at the CDAW Data Center by NASA and The Catholic University of America in cooperation with the Naval Research Laboratory. SOHO is a project of international cooperation between ESA and NASA. Wilcox Solar Observatory data used in this study was obtained via the web site http://wso.stanford.edu at courtesy of J.T. Hoeksema. The Wilcox Solar Observatory is currently supported by NASA. ML acknowledges that this publication is part of the I+D+i project PID2023-147708NB-I00 funded by MICIU/AEI/10.13039/501100011033/ and by FEDER, EU.

%%%%%%%%%%%%%%%%%%%%%%%%%%%%%%%%%%%%%%%%%%%%%%%%%%
\section*{Data Availability}

%The inclusion of a Data Availability Statement is a requirement for articles published in MNRAS. Data Availability Statements provide a standardised format for readers to understand the availability of data underlying the research results described in the article. The statement may refer to original data generated in the course of the study or to third-party data analysed in the article. The statement should describe and provide means of access, where possible, by linking to the data or providing the required accession numbers for the relevant databases or DOIs.

The data and code used to obtain the results of this paper can be available at:
\href{https://github.com/miguelrq21-spec/Long-period-longitudinal-oscillations-in-a-solar-filament-beyond-the-theoretical-period-limit.git}{https://github.com/miguelrq21-spec}.\\
In this paper, we used the SunPy open-source software package \citep{sunpy_community2020}



\bibliographystyle{mnras}
\bibliography{example} % if your bibtex file is called example.bib

%\begin{thebibliography}{99}
%\end{thebibliography}

%%%%%%%%%%%%%%%%%%%%%%%%%%%%%%%%%%%%%%%%%%%%%%%%%%

%%%%%%%%%%%%%%%%% APPENDICES %%%%%%%%%%%%%%%%%%%%%

%\appendix


%If you want to present additional material which would interrupt the flow of the main paper, it can be placed in an Appendix which appears after the list of references.

%%%%%%%%%%%%%%%%%%%%%%%%%%%%%%%%%%%%%%%%%%%%%%%%%%

% Don't change these lines
\bsp	% typesetting comment
\label{lastpage}
\end{document}